\documentclass[aps,prl,twocolumn,superscriptaddress,preprintnumbers,nofootinbib]{revtex4}
\usepackage{epsfig,psfrag,epstopdf}
\pdfoutput=1
\usepackage{amssymb}
\usepackage{amsmath}
\usepackage{color}
\usepackage{graphicx}
\usepackage{hyperref}
\usepackage{chngcntr}
\usepackage{diagbox}
\usepackage{wasysym}
\usepackage{ulem}
\usepackage{multirow}
\graphicspath{{figures/}}

\def\as{\alpha_s}

\begin{document}
\title{Higgs boson pair production via gluon fusion at N$^3$LO in  QCD  }
\author{Long-Bin Chen}
\email{chenlb@gzhu.edu.cn }
\affiliation{School of Physics and Electronic Engineering, Guangzhou University, Guangzhou 510006, China}
\author{Hai Tao Li}
\email{haitaoli@lanl.gov}
\affiliation{Theoretical Division, Los Alamos National Laboratory, Los Alamos, NM, 87545, USA}
\author{Hua-Sheng Shao}
\email{huasheng.shao@lpthe.jussieu.fr}
\affiliation{Laboratoire de Physique Th\'eorique et Hautes Energies (LPTHE), UMR 7589, Sorbonne Universit\'e et CNRS, 4 place Jussieu, 75252 Paris Cedex 05, France}
\author{Jian Wang}
\email{j.wang@sdu.edu.cn}
\affiliation{School of Physics, Shandong University, Jinan, Shandong 250100, China}

\begin{abstract}
    We present next-to-next-to-next-to-leading order (N$^3$LO) QCD predictions for the Higgs boson pair production via  gluon fusion at hadron colliders in the infinite top-quark mass limit. Besides the inclusive total cross sections at various collision energies, we also provide the invariant mass distribution of the Higgs boson pair. Our results show that the N$^3$LO QCD corrections enhance the next-to-next-to-leading order cross section by  $3.0\%$ ($2.7\%$) at  $\sqrt{s}=13~(100)$ TeV, while the scale uncertainty is reduced substantially  below $3\%$ ($2\%$). 
We also find that a judicious scale choice can significantly improve the perturbative convergence.
For the invariant mass distribution, our calculation demonstrates that  
the N$^3$LO corrections improve the scale dependence but almost do not change the shape.     
\end{abstract}

\maketitle

\section{Introduction}

 The discovery of the Higgs boson~\cite{Aad:2012tfa,Chatrchyan:2012xdj} marks the completion of the
standard model (SM) of particle physics and the start of a new era for the physics studies at the LHC.
The next primary goal of the LHC is to precisely pin down its interactions with other SM particles 
or itself. In particular, the precision study of the Higgs potential is ultimately crucial for understanding the  electroweak symmetry breaking mechanism. At the LHC, it has been found that its interaction couplings with massive gauge bosons and fermions agree with the 
SM expectations~\cite{Aad:2015gba, Khachatryan:2016vau,Sirunyan:2018koj,Aad:2019mbh},
while there are only quite weak constraints on the trilinear Higgs self-coupling \cite{Sirunyan:2018two,Aad:2019uzh}.
 However, in the future, the experimental probe will be significantly improved as the increase of the integrated luminosity and the collision energy~\cite{Cepeda:2019klc,Contino:2016spe}, and/or by employing novel analyzing methods~\cite{Kim:2018cxf}. Theoretically, there indeed exist beyond-the-SM (BSM) models, in which  the trilinear Higgs self-coupling deviates from the 
SM value by about $100\%$ while the Higgs couplings to gauge bosons and fermions are almost SM-like~\cite{Kanemura:2002vm}. 
Therefore, the precise measurement of the trilinear Higgs self-coupling at the LHC and future high-energy colliders would be of paramount importance 
to  test the SM
and to  explore the elusive BSM signals.

The direct manner to probe the trilinear Higgs self-coupling at a hadron collider, such as the LHC, is via the Higgs boson pair production. 
Like the single Higgs case, the gluon-gluon fusion (ggF) channel is dominant, while other channels like vector-boson fusion (VBF) are at least one order of magnitude lower in their yields~\cite{Baglio:2012np,Frederix:2014hta}. Similarly to the cross section of single Higgs boson production, the ggF di-Higgs cross section is plagued with large theoretical uncertainties, dominated by the QCD scale uncertainty~\cite{Dawson:1998py} and the top-quark mass scheme dependence~\cite{Plehn:1996wb,Baglio:2018lrj}. The computations of the cross section have been carried out both in the infinite top-quark mass limit and with full top-quark mass dependence. 

In the infinite top-quark mass limit, the next-to-leading order (NLO) QCD correction  was known twenty years ago~\cite{Dawson:1998py}, 
and the NNLO QCD calculation was performed recently~\cite{deFlorian:2013uza, deFlorian:2013jea,Grigo:2014jma,deFlorian:2016uhr}.
In addition, the effect of soft gluon resummation has been investigated at next-to-next-to-leading logarithmic accuracy~\cite{Shao:2013bz,deFlorian:2015moa,deFlorian:2018tah}.

On the other hand, there are also many attempts to go beyond the infinite top-quark mass approximation.
The full top-quark mass dependence was first included in the real-emission part at NLO~\cite{Frederix:2014hta,Maltoni:2014eza}.
The NLO virtual corrections, involving multi-scale two-loop integrals since the LO is already a loop-induced process, have been evaluated by expansion in the heavy top-quark mass limit up to $\mathcal{O}(1/m_t^{12})$ ~\cite{Grigo:2013rya,Grigo:2015dia,Degrassi:2016vss}, in the small top-quark mass limit~\cite{Davies:2018ood,Davies:2018qvx}, and in terms of a small Higgs transverse momentum~\cite{Bonciani:2018omm} or a small Higgs mass~\cite{Xu:2018eos}.
Recently, the expansion of the three-loop virtual corrections in the heavy top-quark mass limit has been presented \cite{Davies:2019djw}.
Finally, the full NLO QCD corrections including exact dependence on the top-quark mass were
computed numerically by two groups~\cite{Borowka:2016ehy,Borowka:2016ypz,Davies:2019dfy,Baglio:2018lrj}, either by using a quasi-Monte Carlo method~\cite{Li:2015foa,Dick2004493} or via a direct Monte Carlo integration by {\sc\small Vegas}.
The matching to parton showers has also been  carried out~\cite{Heinrich:2017kxx,Jones:2017giv,Heinrich:2019bkc}.

Although the infinite top-quark mass approximation is usually insufficient for the corresponding phenomenology studies, a standard way of improving the theoretical prediction on the ggF di-Higgs cross section is to use the lower-order result with full top-quark mass dependence, and to augment it with higher-order corrections in the infinite top-quark mass limit~\cite{Grazzini:2018bsd}.

In this Letter, we provide the first next-to-next-to-next-to-leading order (N$^3$LO) perturbative QCD predictions for the Higgs boson pair production via gluon fusion at hadron colliders in the infinite top-quark mass limit. 
This result becomes one of a few highest-precision computations for scattering processes relevant at the LHC.
The existing calculations performed at N$^3$LO include the inclusive cross sections of the ggF~\cite{Anastasiou:2015ema,Mistlberger:2018etf}, 
VBF~\cite{Dreyer:2016oyx} and bottom-quark fusion~\cite{Duhr:2019kwi} of single Higgs boson production, as well as the VBF of di-Higgs production~\cite{Dreyer:2018qbw}. 
Some differential distributions approximated at the same order are also known for the ggF of single Higgs boson production~\cite{Dulat:2017prg,Cieri:2018oms,Dulat:2018bfe}. 
In our calculation, we will provide both the inclusive cross sections and the invariant-mass distribution of the Higgs boson pair at N$^3$LO, where the latter is the first exact N$^3$LO differential distribution for the ggF channel.
After the completion of this work, the N$^3$LO corrections to the Drell-Yan processes 
are presented  in \cite{Duhr:2020seh}, where the invariant mass distribution of the lepton pair is obtained at N$^3$LO.

\section{Theoretical framework}

In the infinite top-quark mass limit, the effective Lagrangian describing the coupling of two gluon field strength tensors with one or two Higgs bosons  
reads
\begin{align}\label{eq:effL}
 \mathcal{L}_{\rm eff}= -\frac{1}{4}  G_{\mu\nu}^a G^{a~\mu\nu}
 \left(
  C_{h} \frac{h}{v} - C_{hh} \frac{h^2}{2v^2}  
 \right)\,,
\end{align}
where the vacuum expectation value of the Higgs field $v$ can be related to the Fermi constant by $v=\left(\sqrt{2}G_F\right)^{-1/2}$.  
 $C_h$ and $C_{hh}$ are the Wilson coefficients by matching the full theory to the effective theory, 
 which start from $\mathcal{O}(\as)$ and have been calculated up to $\mathcal{O}(\alpha_s^4)$~\cite{Inami:1982xt, Chetyrkin:1997iv, Chetyrkin:1997un, Schroder:2005hy, Chetyrkin:2005ia, Kniehl:2006bg,Baikov:2016tgj,Spira:2016zna,Gerlach:2018hen}.  

\begin{figure}[hbt!]
    \centering
    \includegraphics[scale=0.22]{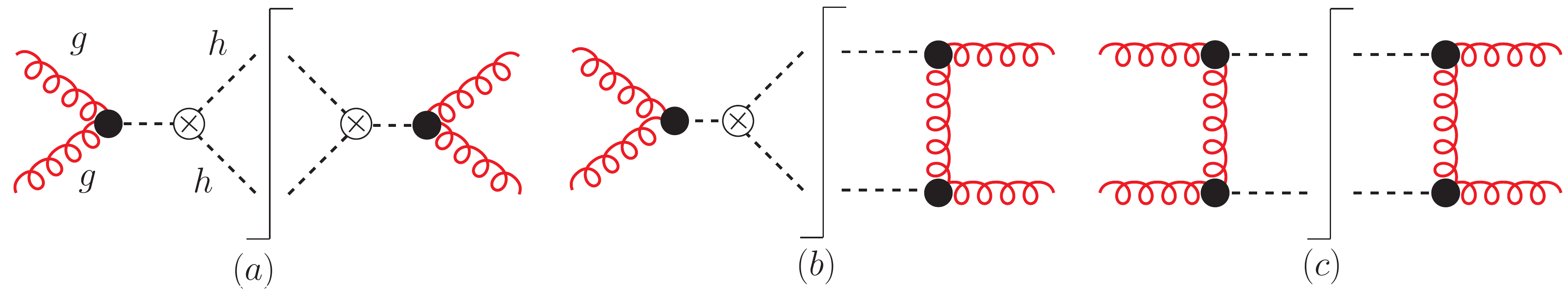}
    \vspace{-0.2cm}
    \caption{Representative Born cut-diagrams for the Higgs boson pair production in ggF. 
    The bullets denote the vertices described by the effective Lagrangian in Eq.(\ref{eq:effL})
    and the crossed circle represents the trilinear Higgs self-coupling. }
    \label{fig:FeynDia}
\end{figure}

The ggF di-Higgs cross section can be organized according to the number of the effective vertex insertions in the amplitude squared, where three representative Born cut-diagrams are shown in Fig.~\ref{fig:FeynDia}. They are the ones with two, three and four effective vertices, and are denoted  as class-$a$, -$b$ and -$c$ respectively in the following context. 
Accordingly, the (differential) cross section can be split into three parts, 
\begin{align}
    d\sigma_{hh} = d\sigma^{a}_{hh} + d\sigma^{b}_{hh} + d\sigma^{c}_{hh} .
\end{align}
Their contributions to various $\alpha_s$ orders are tabulated in Table~\ref{tab:my_label}. At N$^3$LO in $\alpha_s$, we need to calculate the class-$a$ (class-$b$ and class-$c$) contribution at $\rm N^3LO_a$ ($\rm NNLO_b$ and $\rm NLO_c$).
Here the subscripts denote the perturbative orders in the specified class.

\begin{table}[hbt!]
    \centering
    \begin{tabular}{c|c|c|c|c}
    \hline \hline 
        &  LO  & NLO  & NNLO  & N$^3$LO  \\
    \hline
    Total  &  $\mathcal{O}(\alpha_s^2)$  &   $\mathcal{O}(\alpha_s^3)$ &   $\mathcal{O}(\alpha_s^4)$  & $\mathcal{O}(\alpha_s^5)$
     \\
    \hline
    \multirow{2}{*}{a}
     &  ${\rm LO_a} $  &   ${\rm NLO_a}$ &   ${\rm NNLO_a}$  & ${\rm N^3LO_a}$
     \\
      &  $\mathcal{O}(\alpha_s^2)$  &   $\mathcal{O}(\alpha_s^3)$ &   $\mathcal{O}(\alpha_s^4)$  & $\mathcal{O}(\alpha_s^5)$
     \\
    \hline
     \multirow{2}{*}{b}
     &  --- &   ${\rm LO_b}$ &   ${\rm NLO_b}$  &  ${\rm NNLO_b}$
     \\
     &  0 &   $\mathcal{O}(\alpha_s^3)$ &   $\mathcal{O}(\alpha_s^4)$  & $\mathcal{O}(\alpha_s^5)$
     \\     
    \hline
    \multirow{2}{*}{c}
      & --- &  --- &   ${\rm LO_c}$  & ${\rm NLO_c}$ 
     \\
       & 0 &  0 &   $\mathcal{O}(\alpha_s^4)$  & $\mathcal{O}(\alpha_s^5)$
     \\    
     \hline \hline 
    \end{tabular}
    \caption{The perturbative orders in $\alpha_s$ for different classes at the amplitude squared level.
    The first and second rows show the perturbative expansion orders of the cross section of Higgs pair production,
    while the individual class possesses its own expansion orders that are specified  by the subscript.
    For example, we call the $\mathcal{O}(\alpha_s^3)$ contribution in class-$b$ as the ${\rm LO_b}$ in this class since it is the first nonvanishing term,
    though it is an  NLO correction to the cross section of Higgs pair production. }
    \label{tab:my_label}
\end{table}

Because of the similar topologies, the class-$a$ part can be obtained from the calculation
of a single Higgs boson production. They are related by
\begin{align}
    \frac{d\sigma^{a}_{hh}}{dm_{hh}} = f_{h\to hh} 
    \bigg(\frac{C_{hh}}{C_h} - \frac{6 \lambda v^2}{m_{hh}^2-m_h^2} \bigg)^2 
    \times \sigma_{h} (m_h\to m_{hh}),
    \label{eq:htohh}
\end{align}
where $\lambda$ is the Higgs self-coupling and
 the function $f_{h\to hh}$ accounts for the phase space difference between the single and double Higgs boson production,  
\begin{align}
    f_{h\to hh} = \frac{\sqrt{m_{hh}^2- 4 m_h^2}}{16 \pi^2 v^2} ,
\end{align}
and $\sigma_{h} (m_h\to m_{hh})$ denotes the cross section calculated using {\sc\small iHixs2}~\cite{Dulat:2018rbf}
after replacing the Higgs boson mass with the invariant mass of the Higgs boson pair in the code. 
Such a method has also been used in the earlier $\rm NNLO_a$ calculation of the ggF di-Higgs production in Ref.~\cite{deFlorian:2013jea}.

The  class-$b$ part can be obtained through the $q_T$-subtraction method~\cite{Catani:2007vq}, in which we divide the cross section into two parts,
\begin{align}
    d\sigma_{hh}^{b} = d\sigma_{hh}^{b}\Big|_{p_T^{hh}<p_T^{\rm veto}} + d\sigma_{hh}^{b}\Big|_{p_T^{hh}>p_T^{\rm veto}},
    \label{eq:cutoff}
\end{align}
where $p_T^{hh}$ represents the transverse momentum of the Higgs  pair system.
An artificial cutoff parameter $p_T^{\rm veto} $ is introduced in order to deal with the infrared divergences.
In the first part, the transverse momentum of the Higgs pair system is required to be less than the cutoff parameter. 
At higher orders, there can be additional soft emissions or collinear emissions along  the beam line in the final state with a small transverse momentum.
They cause complex infrared divergences, which will cancel against those in the virtual corrections at the end.
Instead of calculating them directly, we employ the result in transverse momentum resummation formula.
With a sufficiently small $p_T^{\rm veto}$, safely ignoring all the power-suppressed terms in $p_T^{\rm veto}$,
the cross section admits a factorization form that can resum the large logarithms $\ln^n  (m_{hh}/p_T^{\rm veto} )$ to all orders in $\alpha_s$. Making use of the soft-collinear effective theory~\cite{Bauer:2000ew,Bauer:2000yr,Bauer:2001ct,Bauer:2001yt,Beneke:2002ph}, we
write the factorized cross section as a convolution of the transverse momentum dependent (TMD) beam function, 
soft function and hard function~\cite{Becher:2012yn}. The rapidity divergences appearing in the calculation of the TMD beam function and the soft function need an additional regulator besides the dimensional regularization \cite{Chiu:2011qc,Chiu:2012ir,Becher:2011dz,Li:2016axz}. However, the final physical cross section is independent of such a regulator. The two-loop analytical results for these ingredients can be found  in~\cite{Gehrmann:2012ze, Gehrmann:2014yya,Luebbert:2016itl,Echevarria:2016scs,Luo:2019hmp}. 
The NNLO hard function can be obtained by combining the two-loop amplitudes calculated in~\cite{Banerjee:2018lfq} and one-loop amplitudes we calculate analytically,
and have been expressed in terms of multiple polylogarithms, which can be evaluated by the public {\sc\small Mathematica} package {\sc\small PolyLogTools}~\cite{Duhr:2019tlz}. 
Consequently, the first term on the right hand of Eq.(\ref{eq:cutoff}) is obtained by expansion of the resummation formula to a fixed order of $\as$. 
We have set up a streamline to combine the various components together in the 
computations of the NNLO differential cross sections of $Whh$~\cite{Li:2016nrr} and $Zhh$~\cite{Li:2017lbf} associated production processes. As opposed to the quark anti-quark initial states in the previous calculations, we extend our program to the gluon-gluon initial states in this work.

In the second part of class-$b$ in Eq.(\ref{eq:cutoff}), the transverse momentum of the Higgs pair system is imposed to be larger than the cutoff parameter $p_T^{\rm veto}$. In such a case, there must be an additional jet accompanying with the Higgs pair. Therefore, in order to have $\rm NNLO_b$ cross section of class-$b$, we only need to calculate the NLO corrections to $hh$ plus a jet,
of which the  underlying Born is represented for example by Fig.\ref{fig:FeynDia}(b) but with an additional gluon emission. 
In this work, we use the {\sc\small MadGraph5\_aMC@NLO}~\cite{Alwall:2014hca} framework to perform such calculations.
The two Wilson coefficients are also expanded in a series of $\alpha_s$.
Since the contribution of this class is from the interference between the amplitudes with only one effective vertex insertion and with two effective vertices, one has to organize these coefficients and amplitudes in 
an appropriate way. 
We have applied, for the first time, the framework in~\cite{Frederix:2018nkq} that 
was originally proposed  to handle mixed-coupling scenarios to our case with a single coupling, $\alpha_s$, but with higher dimensional effective operators, and obtained the results order-by-order  in $\alpha_s$.
To calculate the one-loop amplitudes automatically, 
we construct the model files by using {\sc\small FeynRules}~\cite{Alloul:2013bka}, {\sc\small FeynArts}~\cite{Hahn:2000kx} 
and an in-house {\sc\small Mathematica} program, which has been validated in~\cite{Shao:2011tg,Shao:2012ja}.
We have derived the counter-terms in this model, especially the rational $R_2$ terms, which are needed for automatic computation of the virtual corrections.
We have checked part of them extensively with
the results in the literature~\cite{Draggiotis:2009yb,Page:2013xla}, while
the other part for the vertices with two effective operators are new.
As a consequence, the tensor integrals appearing in the one-loop amplitudes, 
which contain five-point rank-seven integrals,
can then be evaluated by 
{\sc\small MadLoop}~\cite{Hirschi:2011pa,Alwall:2014hca} equipped with {\sc\small Collier}~\cite{Denner:2016kdg}, while the real emission contribution is computed with the module {\sc\small MadFKS}~\cite{Frederix:2009yq,Frederix:2016rdc} with the FKS subtraction method~\cite{Frixione:1995ms,Frixione:1997np}. We want to stress that the inclusion of the contribution from class-$b$ is indispensable in the sense that it not only contributes to the same order in $\alpha_s$ but also cancels the remaining scale dependence in class-$a$ at N$^3$LO; see the appendix.

Finally, since the NLO cross sections of class-$c$ can be obtained with full-fledged methods, we refrain ourselves from presenting details about them, but they have been routinely included in our final results.

\begin{figure}[hbt!]
\centering
\includegraphics[width=.85\columnwidth]{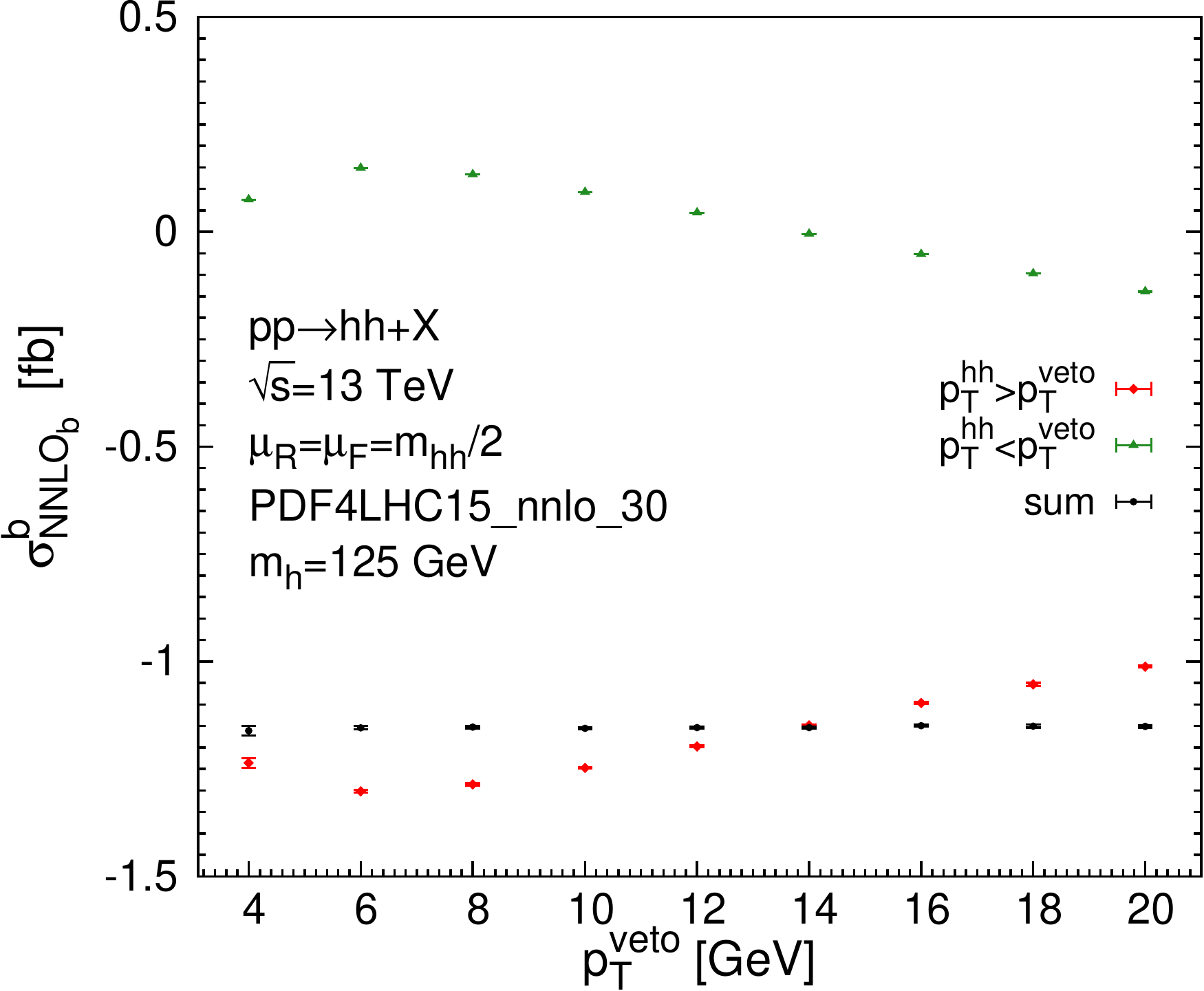}
\caption{The $p_T^{\rm veto}$ dependence of the $\rm NNLO_b$ cross section for class-$b$ at the 13 TeV LHC. The error bars denote the Monte Carlo integration uncertainties.}
\label{fig:ptveto}
\end{figure}

We have performed many cross checks and validations in our calculations. All the terms except for $\mathcal{O}(\alpha_s^5)$ terms of class-$a$ and class-$b$ listed in Table~\ref{tab:my_label} have been cross checked by at least two independent calculations at the inclusive total cross section level. Specifically, we have reproduced the cross section of a single  Higgs boson production up to NNLO in {\sc\small iHixs2} by using our program. This agreement can check our implementations of the two-loop beam and soft functions, as well as the calculation of one-loop amplitudes with one effective vertex. In addition, we have calculated the NLO and NNLO corrections to Higgs pair production in the infinite top-quark mass limit, and found agreement with {\sc\small HPair2}~\cite{Plehn:1996wb,Dawson:1998py} and Ref.\cite{deFlorian:2016uhr}, respectively. This helps to check Eq.(\ref{eq:htohh}) and our calculation of  one-loop amplitudes with two effective vertices. These nontrivial checks already ensure the correctness of many components of our calculations. For the $\mathcal{O}(\alpha_s^5)$ term of class-$a$, we simply used {\sc\small iHixs2} by employing Eq.(\ref{eq:htohh}). Such a program has been validated with the Higgs pair cross sections from LO to NNLO, which makes us convinced that the $\mathcal{O}(\alpha_s^5)$ piece of class-$a$ is correct. For the remaining $\mathcal{O}(\alpha_s^5)$ part of class-$b$, we carefully checked the various pieces that are used in our calculation.  In particular, we have checked the scale dependence of the finite part in the two-loop amplitudes with two effective vertices~\cite{Banerjee:2018lfq} by the renormalization group equation that the hard function should satisfy. 
The one-loop amplitude can also been extracted from the scale-dependent part of the two-loop amplitudes, and it has been compared against the analytical result we calculated with the assistance of {\sc\small fire}~\cite{Smirnov:2014hma} and to the numerical result from {\sc\small MadLoop}. Again, we find perfect agreements. Moreover,
we have checked the independence  of the final $\rm NNLO_b$ results for class-$b$ on the values of $p_T^{\rm veto}$
over the range from 4 GeV to 20 GeV,  as shown in Fig.\ref{fig:ptveto}.

\section{Numerical results}

 In our numerical calculations, we take $v=246.2$ GeV and the Higgs boson mass $m_h=125$ GeV. The top-quark pole mass, which enters only into the Wilson coefficients,  is $m_t=173.2$ GeV.  We use the {\tt PDF4LHC15\_nnlo\_30} PDF~\cite{Butterworth:2015oua,Dulat:2015mca,Harland-Lang:2014zoa,Ball:2014uwa} provided by {\sc\small LHAPDF6}~\cite{Buckley:2014ana}, and the associated strong coupling $\as$. The default central scale is chosen to be the invariant mass of the Higgs pair divided by 2, i.e. $\mu_0=m_{hh}/2$, and the scale uncertainty is evaluated through the 9-point variation of the factorization scale $\mu_F$ and the renormalization scale $\mu_R$ in the form of $\mu_{R,F}=\xi_{R,F}\mu_0$ with $\xi_{R},\xi_{F}\in \{0.5,1,2\}$.    

\begin{table}[h]
\centering
\begin{tabular}{|c|cccc|}
\hline
\diagbox[width=1.3cm]{Order}{$\sqrt{s}$} & $13$ TeV & $14$ TeV & $27$ TeV & $100$ TeV\\\hline
LO & $13.80_{-22\%}^{+31\%}$ & $17.06_{-22\%}^{+31\%}$ & $98.22_{-19\%}^{+26\%}$ & $2015_{-15\%}^{+19\%}$ \\
NLO & $25.81_{-15\%}^{+18\%}$ & $31.89_{-15\%}^{+18\%}$ & $183.0^{+16\%}_{-14\%}$ & $3724_{-11\%}^{+13\%}$\\
NNLO & $30.41^{+5.3\%}_{-7.8\%}$ & $37.55^{+5.2\%}_{-7.6\%}$ & $214.2^{+4.8\%}_{-6.7\%}$ & $4322_{-5.3\%}^{+4.2\%}$ \\
N$^3$LO & $31.31^{+0.66\%}_{-2.8\%}$ & $38.65^{+0.65\%}_{-2.7\%}$ & $220.2^{+0.53\%}_{-2.4\%}$ & $4438^{+0.51\%}_{-1.8\%}$\\\hline
\end{tabular}
\caption{The inclusive total cross sections (in unit of fb) of Higgs boson pair production at different center-of-mass energies from LO to N$^3$LO.
  The quoted relative uncertainties are from the 9-point scale variations $\mu_{R,F}=\xi_{R,F}m_{hh}/2$ with $\xi_{R},\xi_{F}\in \{0.5,1,2\}$. The errors due to the numerical Monte Carlo integration are well below 1\permil.}
\label{tab:totxs}
\end{table}

\begin{figure}[ht]
    \centering
    \includegraphics[width=0.45\textwidth]{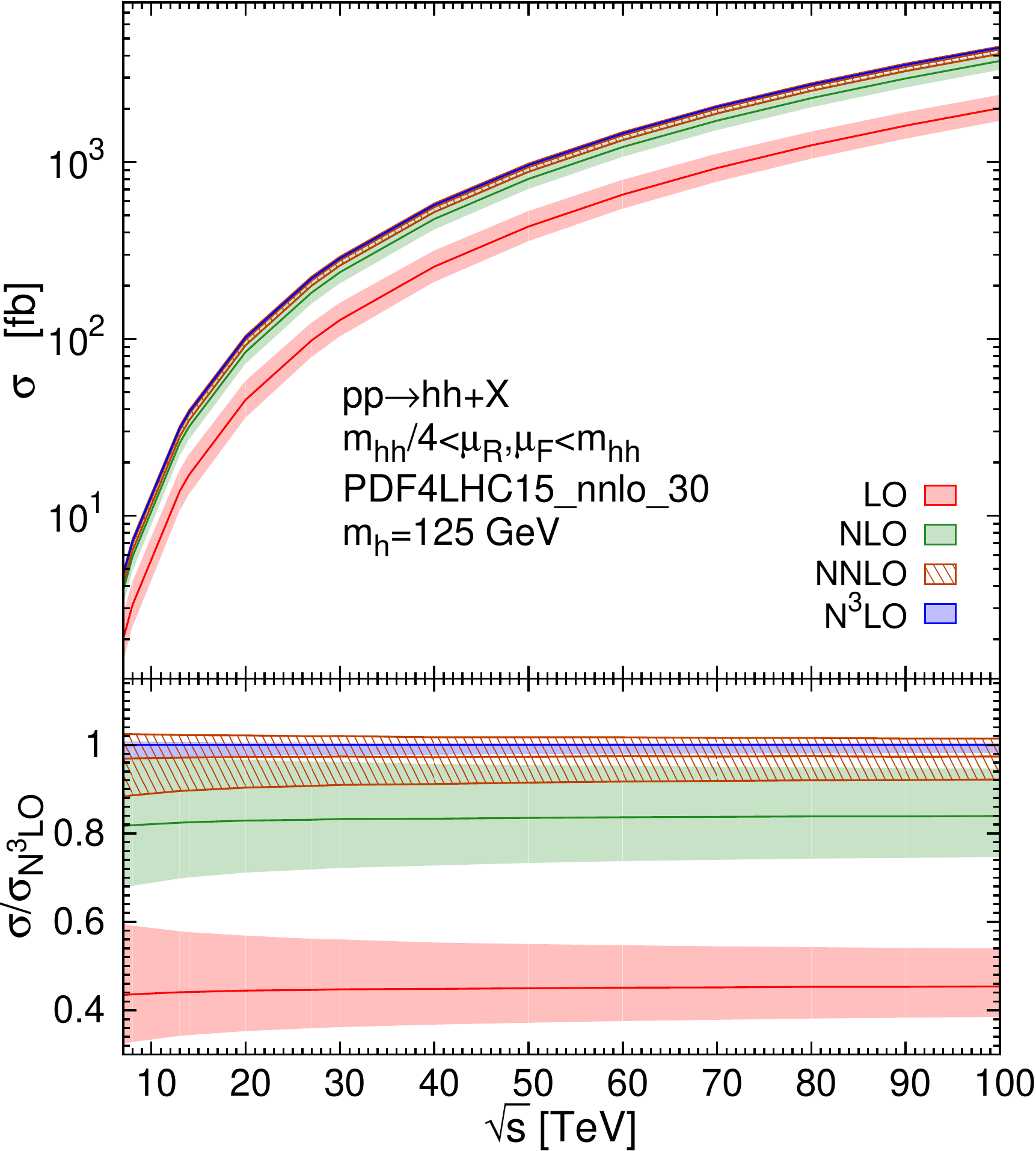}
    \vspace{0cm}
    \caption{The inclusive total cross sections for Higgs boson pair production at proton-proton colliders as a function of the collision energy. 
    The bands represent the scale uncertainties. The bottom panel shows the ratios to the N$^3$LO result.   }
    \label{fig:sigma}
\end{figure}

\begin{figure}[ht]
    \centering
    \includegraphics[width=0.45\textwidth]{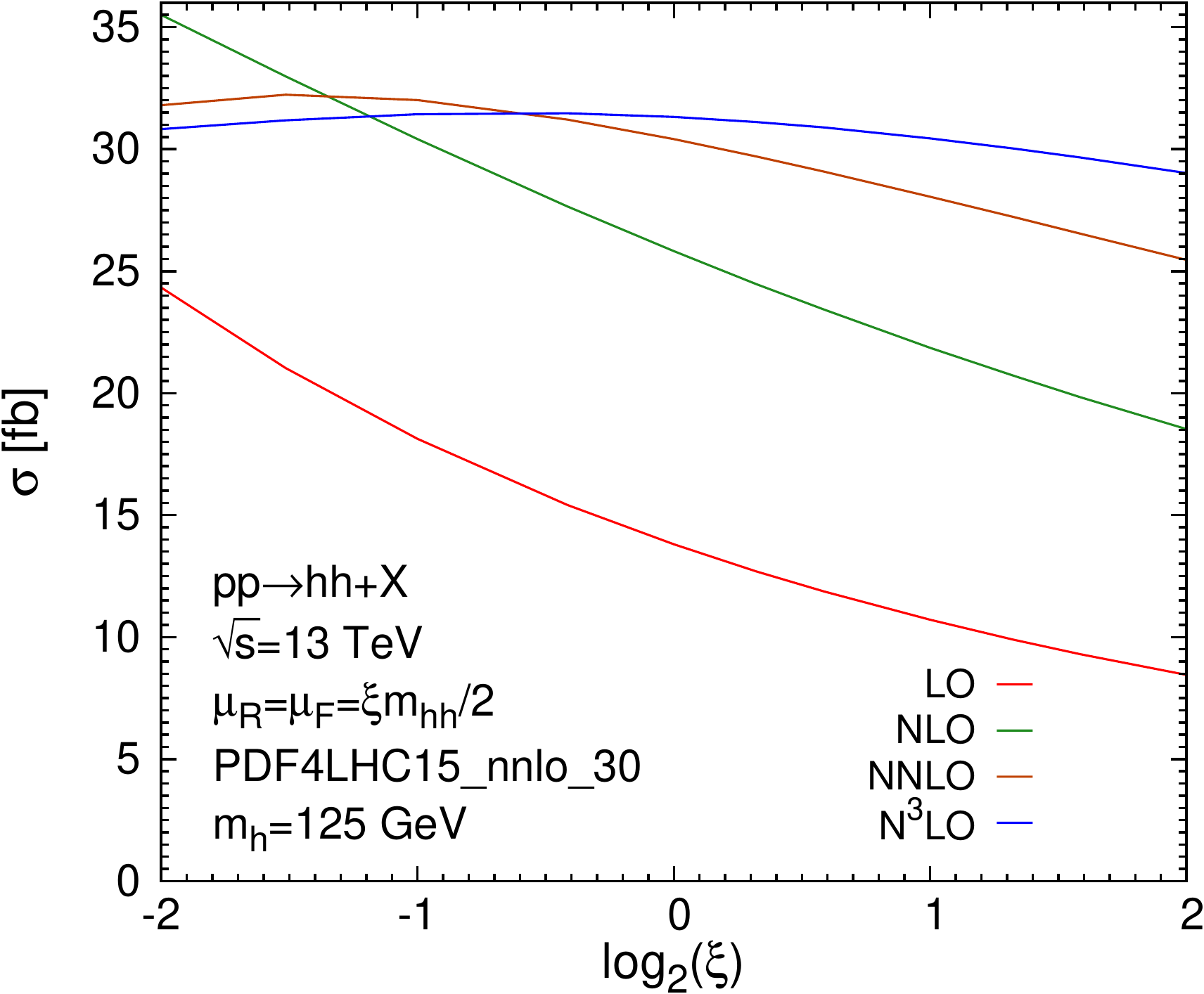}
    \vspace{0cm}
    \caption{The scale dependence of the total cross section for Higgs boson pair production at the LHC with $\sqrt{s}=13$ TeV. 
    We set $\mu_R=\mu_F=\xi \mu_0$ in this plot. }
    \label{fig:scale}
\end{figure}

We present the inclusive total cross sections (from LO to N$^3$LO) of the Higgs boson pair production at different center-of-mass energies in Table~\ref{tab:totxs} and Fig.~\ref{fig:sigma}. Similarly to the single Higgs case, the QCD higher-order corrections are prominent. The NLO corrections increase the LO cross section by  $87\%$  ($85\%$) at $\sqrt{s}=13~(100)$ TeV.
The NNLO corrections increase the NLO cross section further by   $18\%$ ($16\%$), reducing the scale uncertainty by a factor of 2 to 3 to be below $8\%$. Finally, the N$^3$LO corrections turn out to be  $3.0\%$ ($2.7\%$), which lies well within the scale uncertainty band of the NNLO result. Now, the scale uncertainty at N$^3$LO is less than $3\%$ ($2\%$), with another significant reduction of 2-3 times. For the purpose of the comparison, the PDF parameterization uncertainty at 13 TeV amounts to $\pm 3.3\%$, which is larger than the current scale uncertainty. Such an improvement can be more clearly seen in Fig.~\ref{fig:scale}, where we have varied the scale by a factor of four around the default choice with imposing $\mu_R=\mu_F$. The plot illustrates the importance of the choice of scales in a lower order perturbative calculation. If one chooses a scale to be larger than $m_{hh}$, the higher-order QCD corrections are very sizable. Instead, if one chooses a judicious scale between $m_{hh}/4$ and $m_{hh}/3$, 
the perturbative corrections to the inclusive cross section is small from NLO to N$^3$LO.

\begin{figure}[h]
    \centering
    \includegraphics[width=0.45\textwidth]{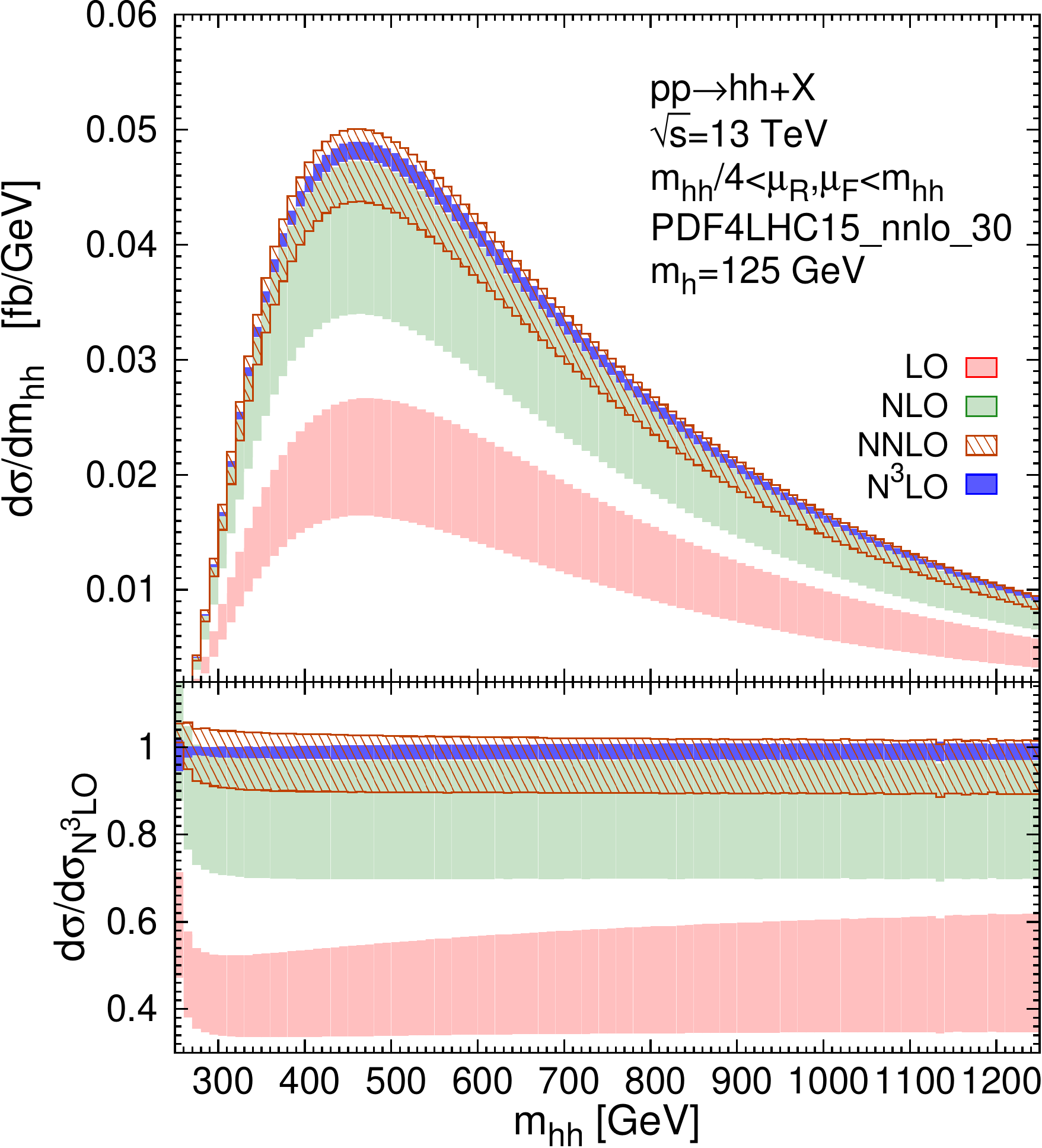}
    \vspace{0cm}
    \caption{Invariant mass distributions for Higgs boson pair production at the LHC with $\sqrt{s}=13$ TeV. The bands represent the scale uncertainties.  The red, green, brown and blue bands correspond to the LO, NLO, NNLO and N$^3$LO predictions, respectively. The bottom panel shows the ratios to the N$^3$LO distribution.  }
    \label{fig:mhh}
\end{figure}

Besides the inclusive total cross section, we are also able to obtain the exact N$^3$LO results for a differential distribution, i.e., the invariant mass $m_{hh}$ distribution shown in Fig.\ref{fig:mhh}. As in the total cross section case, the inclusion of the N$^3$LO corrections dramatically stabilizes the perturbative calculation of the invariant mass differential distribution. It can also be seen that the higher-order QCD corrections do not change the peak position, and the K factor of N$^3$LO over NNLO is almost flat over a large region of $m_{hh}$. The N$^3$LO result with  small scale uncertainty is completely enclosed within the NNLO uncertainty band.
Such a feature consolidates that the perturbative expansion of this differential cross section in a series of $\alpha_s$  is converging up to this order.

\section{Conclusions}

 We have calculated the N$^3$LO QCD corrections to the Higgs boson pair production via gluon fusion
at hadron colliders in the infinite top-quark mass limit.
We find that the total cross section at N$^3$LO increases by $3.0\%$ ($2.7\%$)  at $\sqrt{s}=13~(100)$ TeV with respect to the NNLO result under the central scale choice $\mu_0=m_{hh}/2$. The scale uncertainty has been significantly improved at N$^3$LO compared to the previous result at NNLO, which is now less than $3\%$ ($2\%$). In contrast, the PDF  uncertainty is $\pm 3.3\%$ at the 13 TeV LHC.
Moreover, we have computed the invariant mass distribution at N$^3$LO for the first time, 
and the shape of the distribution is almost unchanged.
The perturbative series of both the total inclusive cross section and the invariant mass distribution are found to be converging up to this order.

In the future, for the phenomenological applications, it is essential to combine our N$^3$LO calculation in the infinite top-quark mass limit with the NLO result including exact top-quark mass dependence \cite{Chen:2019fhs}.
The NNLO result with finite top-quark mass is not available at the moment.
A preliminary investigation indicates that the uncertainty from taking infinite top-quark mass limit at NNLO
can be as large as $\pm 3\%$ ($\pm 5\%$) at 13 (100) TeV~\cite{Grazzini:2018bsd}.
Though this is obtained under certain assumptions, it indicates that the top-quark mass effects are more important at larger collider energies.
It would be interesting and informative to compute the exact NNLO result with full top-quark mass dependence
in order to reduce the theoretical uncertainty further.

\section*{Acknowledgments}

We thank D. Y. Shao for collaborations at the early stage of this work.
We also thank C. Lee and Y.-Q. Ma for carefully reading the manuscript.
LBC was supported by the National Natural Science Foundation of China (NSFC) under the grants 11747051 and 11805042.
HTL was supported by the Los Alamos National Laboratory LDRD program.
The work of HSS was supported by the ILP Labex (ANR-11-IDEX-0004-02, ANR-10-LABX-63). 
JW was  supported  by the BMBF project 05H18WOCA1 when he was in Technische Universit\"at M\"unchen
and by the program for Taishan scholars.

\onecolumngrid

\section*{Appendix: Renormalization scale dependence}

In this appendix,  
we present the method to obtain the $\mu_R$ dependence in the framework of soft-collinear effective theory (SCET), where the renormalization scale is usually set to be the same as the factorization scale in the expansion of the resummed formula,
while they can be distinguished clearly in  fixed-order calculations.


The  cross section is scale $(\mu_R=\mu_F=\mu)$ invariant,
\begin{align}
    \frac{d}{d\ln \mu} \sigma_{hh}(\mu,\mu) = \left(\frac{d}{d\ln \mu_R} \sigma_{hh}(\mu_R, \mu_F)  + \frac{d}{d\ln \mu_F} \sigma_{hh}(\mu_R, \mu_F) \right)\bigg|_{\mu_R=\mu_F=\mu} = 0 + \mathcal{O}(\as^6).   
\end{align}
The individual renormalization and factorization scale dependence can be obtained through 
\begin{align}
    \sigma_{hh}(\mu_R, \mu_F)= \sigma_{hh}(\mu_F, \mu_F) + \int_{\mu_F}^{\mu_R} d\bar{\mu} \left(\frac{d}{d\mu_R} \sigma_{hh}(\mu_R, \mu_F) \bigg|_{\mu_R=\bar{\mu}}\right) \,,
\end{align}
where the first part on the right hand can be predicted with $q_T$-subtraction method in the framework of SCET
and the second part is obtained by requiring the renormalization scale independence of the total cross section.

The N$^3$LO cross section for Higgs pair production is renormalization scale invariant up to $\mathcal{O}(\as^6)$ corrections, i.e. 
\begin{align}
    \frac{d}{d\ln\mu_R} \sigma_{hh} (\mu_R, \mu_F) =  \frac{d}{d\ln\mu_R} \sigma_{hh}^{a}(\mu_R, \mu_F)
    + \frac{d}{d\ln\mu_R} \sigma_{hh}^{b}(\mu_R, \mu_F)+ \frac{d}{d\ln\mu_R} \sigma_{hh}^{c} (\mu_R, \mu_F) = 0 + \mathcal{O}(\as^6)\;.
\end{align}
For class-$a$, the differential equation is 
\begin{align}
    \frac{d}{d\ln \mu_R}\sigma_{hh}^{a} (\mu_R, \mu_F) = \int dm_{hh} f_{h\to hh} [ \sigma_{h} (\mu_R, \mu_F,m_h\to m_{hh})]  \times \frac{d}{d\ln \mu_R}  \bigg(\frac{C_{hh}}{C_h} - \frac{6 \lambda v^2}{m_{hh}^2-m_h^2} \bigg)^2 \;.
\end{align}
where $\sigma_{h}$ has the expansion $\sigma_{h}=\sigma_{h}^{(0)}+\sigma_{h}^{(1)}+\dots$ with $\sigma_{h}^{(i)}\propto \as^{2+i}$.
Up to N$^3$LO, we need the NLO QCD corrections to class-$c$ cross section which is standalone and scale invariant. Therefore, for class-$b$, the renormalization group equation is 
\begin{align} \label{eq:RGb}
     \frac{d}{d\ln \mu_R} \sigma^b_{hh}(\mu_R, \mu_F) = -2\int dm_{hh} f_{h\to hh} [ \sigma_{h} (\mu_R, \mu_F, m_h\to m_{hh})]\times \bigg(\frac{C_{hh}}{C_h} - \frac{6 \lambda v^2}{m_{hh}^2-m_h^2} \bigg)  \left( \frac{d}{d\ln \mu_R}  \frac{C_{hh}}{C_h}\right) \,.
\end{align}
The ratio of $C_{hh}$ over $C_h$ can be written in a series of  $a_s\equiv\as(\mu_R)/4\pi$,
\begin{align}
    \frac{C_{hh}}{C_{h}} = 1+ \delta_2 a_s^2 + \delta_3(\mu_R) a_s^3 + \mathcal{O}(a_s^4) \;.
\end{align}
where the coefficient $\delta_2$ is scale independent.
Therefore, we obtain
\begin{align}
    \frac{d}{d\ln \mu_R}  \frac{C_{hh}}{C_h} =& \left( \frac{da_s}{d\ln\mu_R}\frac{\partial }{\partial a_s } + \frac{\partial }{\partial \ln \mu_R} \right) \frac{C_{hh}}{C_h}
    = -4 \beta_0 \delta_2 a_s^3 +  a_s^3\frac{d \delta_3}{d\ln\mu_R}+ \mathcal{O}(a_s^4)\equiv a_s^3 \eta +  \mathcal{O}(a_s^4)
\end{align}
with $\beta_0=(11 C_A-2 n_f)/3$~. 
Then Eq.~(\ref{eq:RGb}) becomes
\begin{align} \label{eq:RGb2}
     \frac{d}{d\ln \mu_R} \sigma^b_{hh}(\mu_R, \mu_F) =&   -2 a_s^3 \eta\; \int dm_{hh} f_{h\to hh} [ \sigma_{h}^{(0)} (\mu_R, \mu_F,m_h\to m_{hh})]\bigg(1- \frac{6 \lambda v^2}{m_{hh}^2-m_h^2} \bigg)+\mathcal{O}(a_s^6)
     \nonumber \\ 
     =& -\frac{3}{4}a_s^3\; \eta \;  \sigma^{b (1)}_{hh}(\mu_R, \mu_F) +\mathcal{O}(a_s^6),
\end{align}
where the class-$b$ cross section is written as $\sigma^b_{hh} =\sum_{i=1} a_s^i \sigma^{b (i)}_{hh} $ with $\sigma^{b (i)}_{hh} \propto a_s^2$ and 
\begin{align}
    \sigma^{b (1)}_{hh}(\mu_R, \mu_F) =  \frac{8}{3} \int dm_{hh} f_{h\to hh} [ \sigma_{h}^{(0)} (\mu_R, \mu_F,m_h\to m_{hh})]\bigg(1 - \frac{6 \lambda v^2}{m_{hh}^2-m_h^2} \bigg).
\end{align}
The scale-invariance violation terms in Eq.~(\ref{eq:RGb2}) start from NNLO corrections to class-$b$ Higgs pair production.

\bibliography{bib}

\end{document}